# Chirality control via double vortices in asymmetric Co dots

Randy K. Dumas, Dustin A. Gilbert, Nasim Eibagi, and Kai Liu*

*Department of Physics, University of California, Davis, California, 95616*

**Abstract**

Reproducible control of the magnetic vortex state in nanomagnets is of critical importance. We report on chirality control by manipulating the size and/or thickness of asymmetric Co dots. Below a critical diameter and/or thickness, chirality control is achieved by the nucleation of single vortex. Interestingly, above these critical dimensions chirality control is realized by the nucleation and subsequent coalescence of two vortices, resulting in a single vortex with the opposite chirality as found in smaller dots. Micromagnetic simulations and magnetic force microscopy highlight the role of edge-bound halfvortices in facilitating the coalescence process.

**PACS's: 75.70.Kw, 75.60.Jk, 75.75.-c, 75.75.Fk**

Magnetic vortices in sub-micron sized dots have gained considerable interests in recent years due to their unique reversal mechanisms, fascinating topological properties, and potential applications in information storage,[1-6] spin-torque oscillators,[7,8] magnetic memory and logic devices,[9] and targeted cancer-cell destruction strategies.[10] Vortices are one type of topological defects characterized by an in-plane magnetization with a clockwise (CW) or counter-clockwise (CCW) chirality and a central core with an out-of-plane magnetization (up or down polarity). Because chirality and polarity are independent quantities, interesting data storage possibilities arise where a single dot can store two bits of information.[6] Vortex interaction further offers a synchronization route to achieve nanosized spin-torque oscillators for microwave generation.[8,11] Additionally, the dimensional crossover from vortices to domain walls (DWs)[12] leads to the occurrence of vortices in DWs, which influences the DW manipulation by a spin-polarized current.[9] Very recently, halfvortices[13] have been theoretically recognized as another important class of elementary topological defects.[14] These are edge defects with half-integer winding numbers ($n=\pm1/2$), as opposed to vortices with integer ones ($n=\pm1$).[15] The vortex-DW crossover illustrates that DWs are just composites of elementary defects.

The ability to *control* the vortex state in magnetic nanostructures is of critical importance. Switching the vortex core polarity has been demonstrated by the appropriate application of time-varying magnetic fields[6,16] or spin polarized currents.[17-19] On the other hand, vortex chirality is degenerate in symmetrical nanomagnets, such as circular dots. Interestingly, asymmetric structures, e.g. nominally circular nanomagnets with a flattened edge[20,21] or triangularly shaped dots[22], make chirality control possible. The broken symmetry leads to a preferred vortex nucleation site and subsequent chirality control. Once chirality control is established the vortex annihilation site can then be manipulated by an appropriate field sweep. The resulting vortex annihilation field sensitively depends on where the vortex is expelled from the dot.[23] To date, the observed chirality control mechanism has been largely based on the nucleation and annihilation of a *single vortex* in each asymmetric nanomagnet. Here, we report a different and opposite chirality control mechanism through the nucleation and coalescence of *double vortices*. For an identical field sweep a vortex with either CW or CCW chirality can be achieved at remanence by altering the diameter and/or thickness of the dot. We find that halfvortices play an important role in facilitating this particular chirality control mechanism.

Arrays of polycrystalline asymmetric Co dots were fabricated on naturally oxidized Si substrates using standard electron beam lithography and lift-off techniques in conjunction with magnetron sputtering. Each array has 1 nm Ta buffer and capping layers. The dots form a square array over a $50 \times 50 \mu m^2$ area, with a center-to-center separation of 1 µm. This spacing ensures interactions between dots are minimal.[3, 24] Dots with Co thicknesses of 45 - 53 nm and diameters of 650-875 nm were studied. The asymmetry is achieved by flattening the top portion of circular dots, as discussed earlier.[24] A scanning electron microscopy (SEM) image is shown in Fig. 1 (upper left inset) of asymmetric dots with a thickness of 53nm and a diameter of 760nm.

Magnetic hysteresis loops were measured at room temperature using the magneto-optical Kerr effect (MOKE) on a Durham Magneto Optics NanoMOKE2 magnetometer.[1] The beam was focused to a 30 µm diameter spot size, capturing the average reversal behavior of $\sim 10^3$ dots. For each array, major hysteresis loops were measured between ±1100 Oe and half loops were measured over 1100 → 0 → 1100 Oe, both with a field spacing of ~1 Oe and a field sweep rate of 11 Hz. Typically $\sim 10^3$ loops were averaged to obtain a single hysteresis curve. The annihilation field along the major and half loops is quantitatively determined from the field at which the magnetization jumps abruptly, i.e., where the *M-H* curve has a maximum slope. Additionally, atomic and magnetic force microscopy (AFM/MFM) images were acquired using an Asylum Research MFP-3D atomic force microscope in standard phase detection mode with low moment CoCr tips. For both MOKE and MFM measurements, the magnetic field was applied in the plane of the dots along the flat edge (positive field pointing to the right, as shown in Fig. 1 upper left inset). The experimental results were also compared with simulations using the OOMMF code.[25] Material parameters suitable for polycrystalline Co were used (saturation magnetization $M_S$=1.4×10$^6$ A/m, exchange stiffness $A$=1.3×10$^{-11}$ J/m, and crystalline anisotropy was neglected).

The reversal behavior of an array of 45 nm thick asymmetric Co dots with a diameter of 760 nm is shown in Fig. 1 (main panel). The major and half loops show two distinct annihilation fields occurring at 840 and 667 Oe, respectively. The different annihilation fields depend on which side of the dot the vortex is annihilated from. In a prior study we found that vortex annihilation from the flat edge occurs in a smaller field than from the rounded edge.[24] Therefore, by simply analyzing the annihilation field, the vortex chirality can be determined. Additionally, as reported earlier by Schneider et al,[20] if the dot is first positively (negatively) saturated to the right (left), a vortex with CCW (CW) chirality is achieved at remanence by the nucleation of a single vortex from the flat edge of the dot. The reversal behavior is directly examined by MFM. The sample is first saturated in a +2 kOe field. As the applied field is decreased from positive saturation [Fig. 1-panel (i)] to zero, a single vortex is nucleated from the flat edge of the dot [Fig. 1-panel (ii)]. As the applied field is then increased back towards positive saturation (along the half loop) the vortex core moves towards and subsequently annihilates from the flat edge [Fig. 1-panel (iii)]. This annihilation mode confirms the CCW vortex chirality at remanence.

Interestingly, the reversal behavior of slightly larger asymmetric dots, 45nm thickness and 810 nm diameter, is strikingly different (Fig. 2). The annihilation field along the half loop (732 Oe) is now significantly larger than that along the major loop (645 Oe), suggesting that the vortex chirality may be opposite to that of the smaller diameter dots shown in Fig. 1. A detailed account of the magnetization reversal processes is revealed by MFM. As the applied field is reduced from positive saturation [Fig. 2-panel (i)] an unusual buckling of the magnetization is observed at an applied field of 340 Oe [Fig. 2-panel (ii)], unlike that discussed in Ref. [20]. This buckling precedes the nucleation of two vortices from the rounded edge of the dot at a field of 260 Oe [Fig. 2-panel (iii)]. However, these two vortices quickly coalesce into a single vortex with *CW* chirality at remanence [Fig. 2-panel (iv)], opposite to that shown in Fig. 1. As the applied field is then increased back towards positive saturation, along the half loop, the single vortex now moves towards the rounded edge [Fig. 2-panel (v)], confirming the CW chirality at remanence. On the contrary, along the major loop the vortex is expelled from the flat edge of the dot, and does so in a smaller annihilation field. It is important to note that in both the 760 nm and 810 nm diameter dots it is always more difficult to expel the vortex from the rounded edge of the dot. Therefore, analysis of the annihilation field behavior along major and half loops is a robust and reproducible technique to determine vortex chirality.

In order to better understand the more complex reversal and chirality control mechanism observed in the 810 nm diameter dots, micromagnetic simulations have been performed. Simulated major and half loops are shown in Fig. 3(a) along with the domain configurations along the half loop. The annihilation field along the half loop is larger than that along the major loop, consistent with the experiment. The simulated domain configurations also closely reproduce the observed MFM images shown in Fig. 2 panels (i)-(v). Reversing from positive saturation, two edge-bound halfvortices[13] (each with a winding number of $n= -1/2$) first appear [Fig. 3(a)-panel (i)]. This is immediately followed by the nucleation of double vortices (each with $n= +1$) with the same CW chirality and opposite polarity [Fig. 3(a)-panel (ii)]. In fact, upon close inspection these halfvortices are visible in the MFM images where the magnetization appears to be pinched near the edge of the dot [Fig. 2-panels (ii)-(iii), highlighted with small arrows]. As the applied field is reduced further the two vortices coalesce into a single vortex with CW chirality [Fig. 3(a)-panel (iii)]. As the applied field is then reversed back to positive saturation, along the half loop, the vortex core annihilates from the rounded edge of the dot, as shown in panel (iv) of Fig. 3(a).

Additionally, a detailed analysis of the vortex coalescence has been carried out by inspecting individual iterations from the OOMMF micromagnetic solver. This captures instantaneous "snapshots" of the evolving magnetic configuration within the dot, instead of equilibrated control point at each applied field as was done in Fig. 3(a). We find that one CW vortex annihilates with the two halfvortices via a complex process involving the creation of another vortex/anti-vortex pair,[26] leaving behind the other CW vortex. The total winding number is preserved during the coalescence process.

The coalescence can also be understood by considering the various energy contributions during the reversal. The Zeeman, demagnetization, and exchange energy densities are plotted in Fig. 3(b) along the decreasing-field branch of the simulated loops near the initial double vortex nucleation ($H=600$ Oe) and subsequent vortex coalescence ($H=400$ Oe) fields. First we consider the initial nucleation of two vortices. It has been shown, despite the additional gain in exchange energy associated with two vortices, that double-vortex nucleation becomes more probable as the volume of the dot gets larger.[11, 27] In order to fully understand this seemingly energetically costly configuration the Zeeman contribution to the total energy must also be considered. The nucleation of two vortices, with the *same chirality*, keeps a large fraction of the spins aligned with the external field, which results in a relatively small (as compared to single vortex nucleation) increase in Zeeman energy at the double vortex nucleation field. As the double vortices nucleate, highlighted with a vertical dashed line in Fig. 3(b), increases in Zeeman (+14 kJ/m$^3$) and exchange energy densities (+3 kJ/m$^3$) are balanced by a comparable decrease in demagnetization energy density (-13 kJ/m$^3$). However, it then becomes energetically unfavorable to maintain two vortices within the dot as the exchange energy rapidly increases [Fig. 3(b) inset]. Since any further increase in exchange energy would lead to a prohibitively high total energy, the two vortices coalescence ($H$=400 Oe) into a single vortex. The resulting increase in Zeeman energy density (+20 kJ/m$^3$) is offset by a drastic drop in exchange (-5 kJ/m$^3$) and demagnetization (-25 kJ/m$^3$) energy densities.

In order to establish where the cross-over in the two different chirality control mechanisms occurs, asymmetric dot arrays with nominal thicknesses of 45nm and diameters ranging between 650 and 875nm were analyzed. The samples were initially positively saturated in a +2 kOe field before imaging the remanent state. To achieve a degree of statistical significance a 10×10 subset of each array was imaged and dots with either CW or CCW chirality were counted. The experimental results (closed squares) were also compared with simulations (open squares) carried out on a single dot, as shown in Fig. 4 where a chirality control parameter is defined as (CCW-CW)/(CCW+CW). Clearly, as the diameter increases the fraction of dots with CCW chirality decreases. The phase boundary between CCW and CW chirality control occurs for dots with a diameter of about 800 nm. As also shown in Fig. 4, a similar comparison between experimental (closed circles) and simulation (open circles) results was conducted for 53nm thick dots. The crossover between CCW and CW chirality control occurs at about 650 nm, which is significantly smaller than that for the 45 nm thick dots. *Therefore, for a given dot* *diameter, the thickness can also be used to tune the vortex chirality at remanence.* For both the 45 and 53 nm thick dots the experimental and simulated data show overall good quantitative agreement. The simulations are of a single dot, which lead to a sharper phase boundary, whereas experimentally small variations in size/shape within the dot arrays lead to the more gradual boundary.

In summary, we have found a fundamentally different chirality control mechanism involving the nucleation and coalescence of double vortices. This is realized by tailoring the diameter and/or thickness of asymmetric dots, where the asymmetry is introduced by flattening the top of a nominally circular disk. For the 45 nm thick Co dots a critical diameter of about 800 nm is found that separates two distinctly different chirality control mechanisms. For diameters smaller than 800 nm, a single vortex is nucleated from the flat edge of the dot and at remanence the moments along the flat edge of the dot lie anti-parallel to the previously saturated state, achieving chirality control. Therefore a dot initially saturated to the right will result in a CCW vortex at remanence. However, a different mechanism is found for dots with diameters larger than 800 nm that involves the initial nucleation and subsequent coalescence of two vortices which leads to (after positive saturation to the right) a vortex with CW chirality at remanence. In order to conserve the winding number the vortex coalescence is mediated by two halfvortices bound to the dot edge. For 53 nm thick dots, the phase boundary between CW and CCW chirality control is found to be smaller, about 650 nm. These results demonstrate possibilities of tuning dot diameter, thickness, and edge details, in addition to dot asymmetry, to control the remanent state vortex chirality.

This work has been supported by NSF (ECCS-0925626, ECCS-0725902, and DMR-1008791) and CITRIS.

**F**igure Captions

**Fig. 1.** (Color online) SEM image (inset, upper left) of Co dots with a horizontal flat edge. The applied field direction is parallel to the flat edge of the dot and positive fields are defined to the right. The measured major and half loops (main panel) of an array of 45 nm thick asymmetric Co dots with a diameter of 760 nm exhibit different vortex annihilation fields. The MFM images at various applied fields along the half loop are shown in panels (i)-(iii) (scale bar corresponds to 300 nm), where the contrast is projected onto a three-dimensional landscape obtained with tapping-mode AFM.

**Fig. 2.** (Color online) Measured major and half loops (main panel) of an array of 45nm thick asymmetric Co dots with a diameter of 810 nm. MFM images at various applied fields along the half loop are shown in panels (i)-(v) (scale bar corresponds to 300 nm), where the contrast is projected onto a three-dimensional landscape obtained with tapping-mode AFM.

**Fig. 3.** (Color) (a) Simulated major and half loops of a single 45 nm thick asymmetric Co dot with a diameter of 810 nm. The domain configurations at various applied fields along the half loop are shown in panels (i)-(iv). The black arrows indicate the in-plane $x$-$y$-components while the colors (red=up, blue=down) indicate the $z$-component of the magnetization. (b) Simulated energy density changes along the decreasing-field branch of the loop near the double vortex nucleation and coalescence fields.

**Fig. 4.** (Color online) Chirality control order parameter, defined as (CCW-CW)/(CCW+CW), for 45 nm (squares) and 53 nm (circles) thick asymmetric dots as a function of dot diameter. The experimental data (solid symbols) is obtained by counting the number of CCW or CW vortex remanent states of a 10×10 subset of each dot array *after positive saturation to the right*. The simulated data (open symbols) is of a single asymmetric dot.

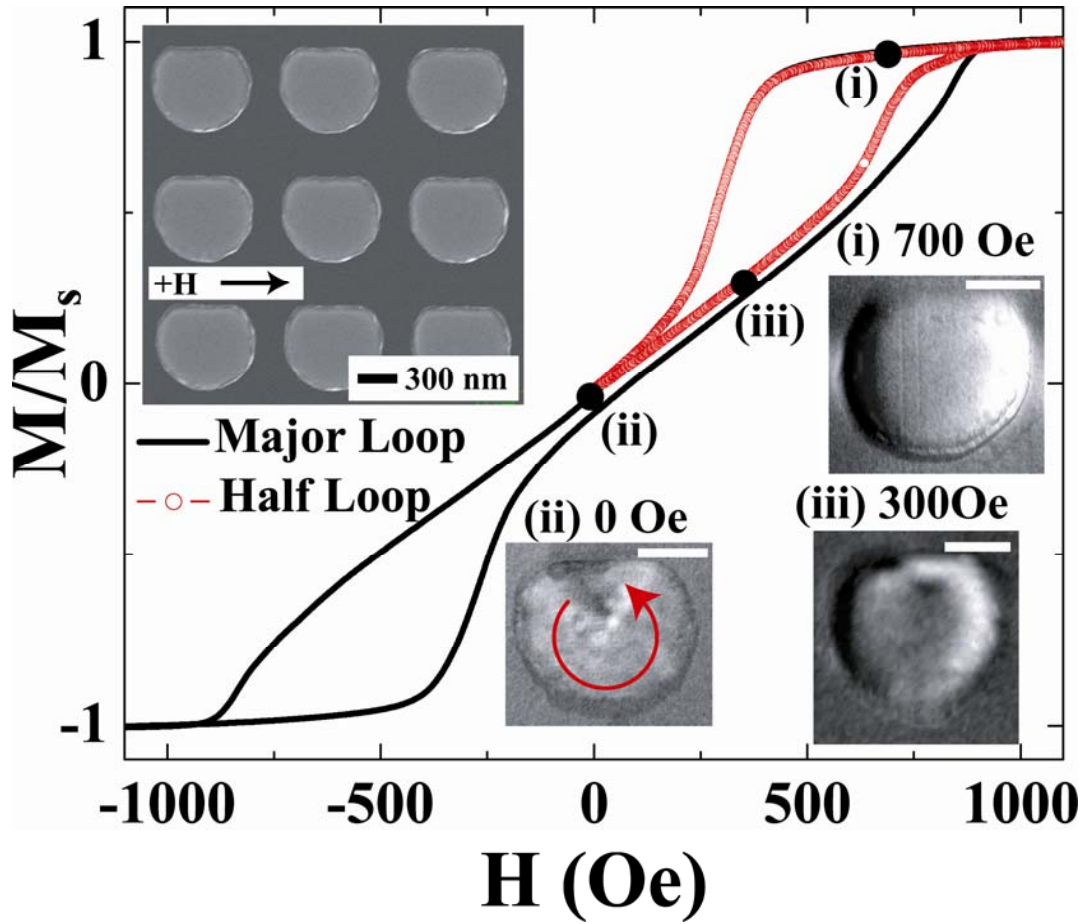

**Fig. 1, Dumas** *et. al.*

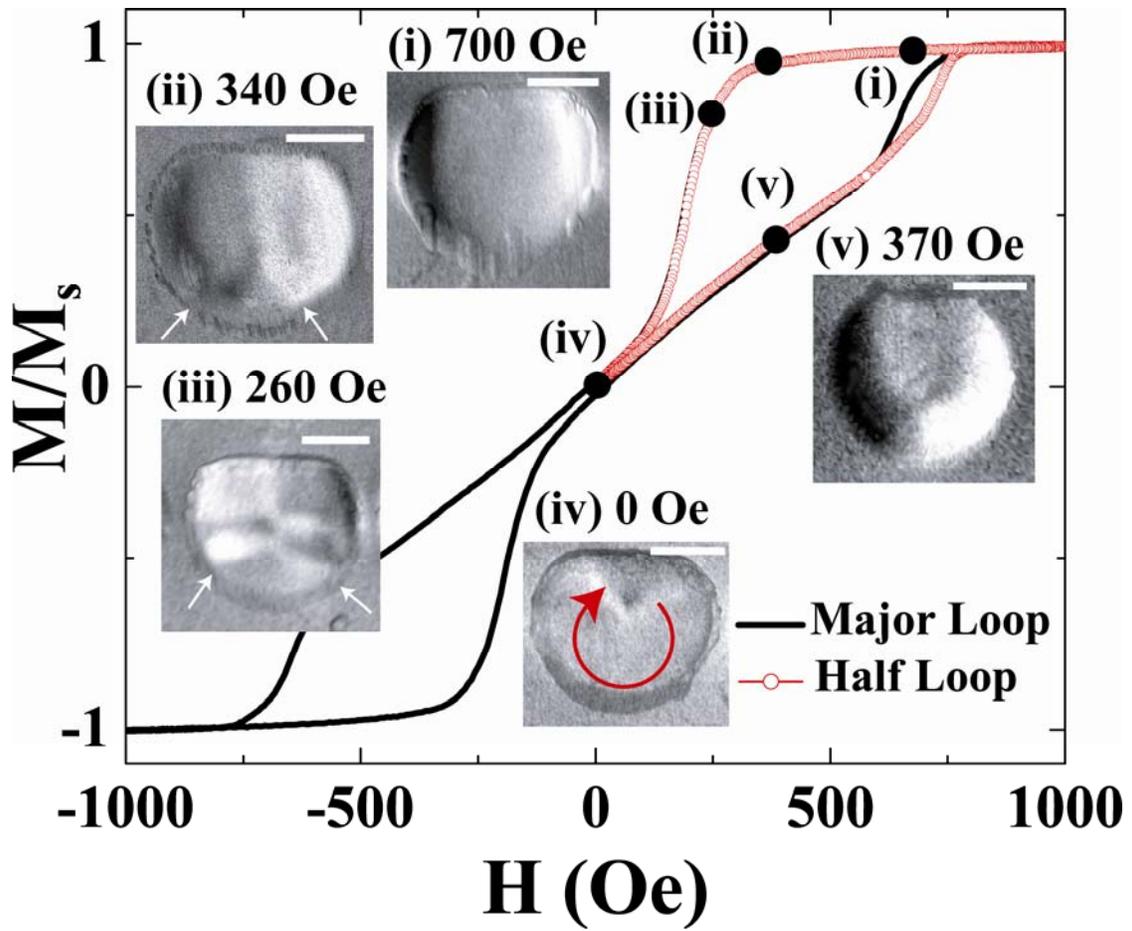

**Fig. 2, Dumas** *et. al.*

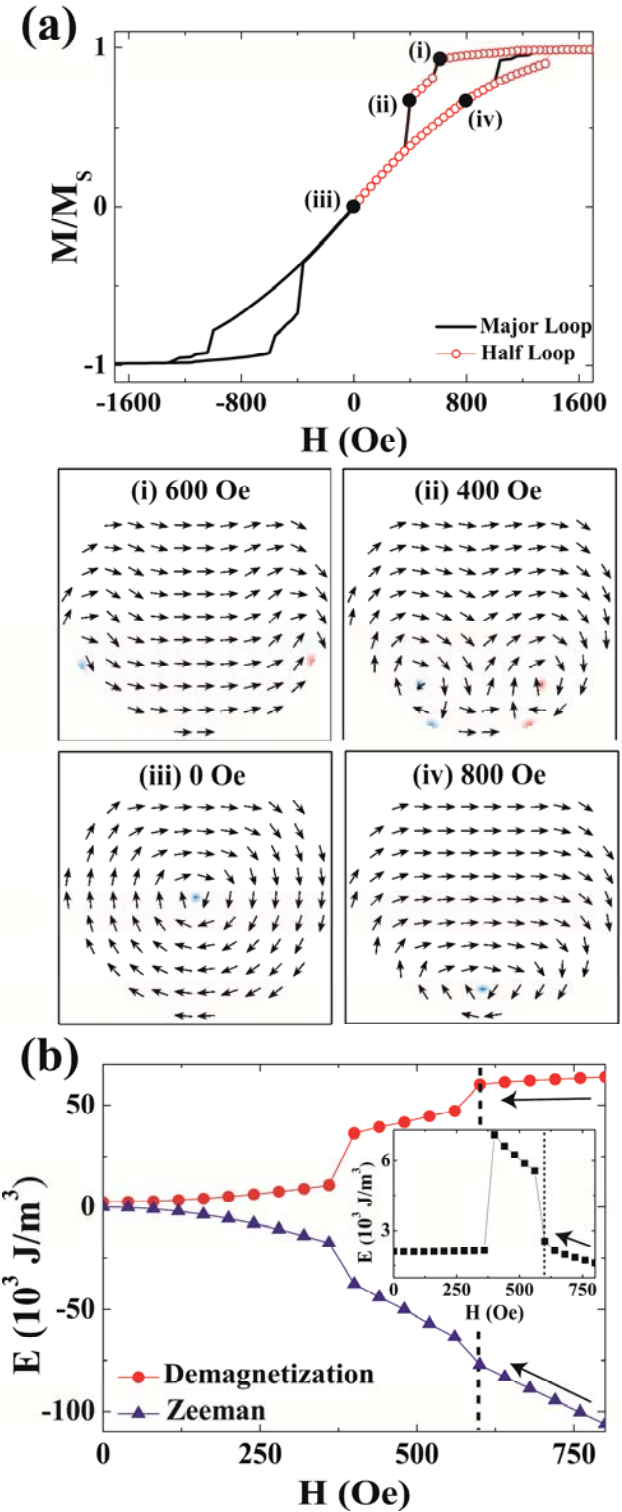

**Fig. 3, Dumas et. al.**

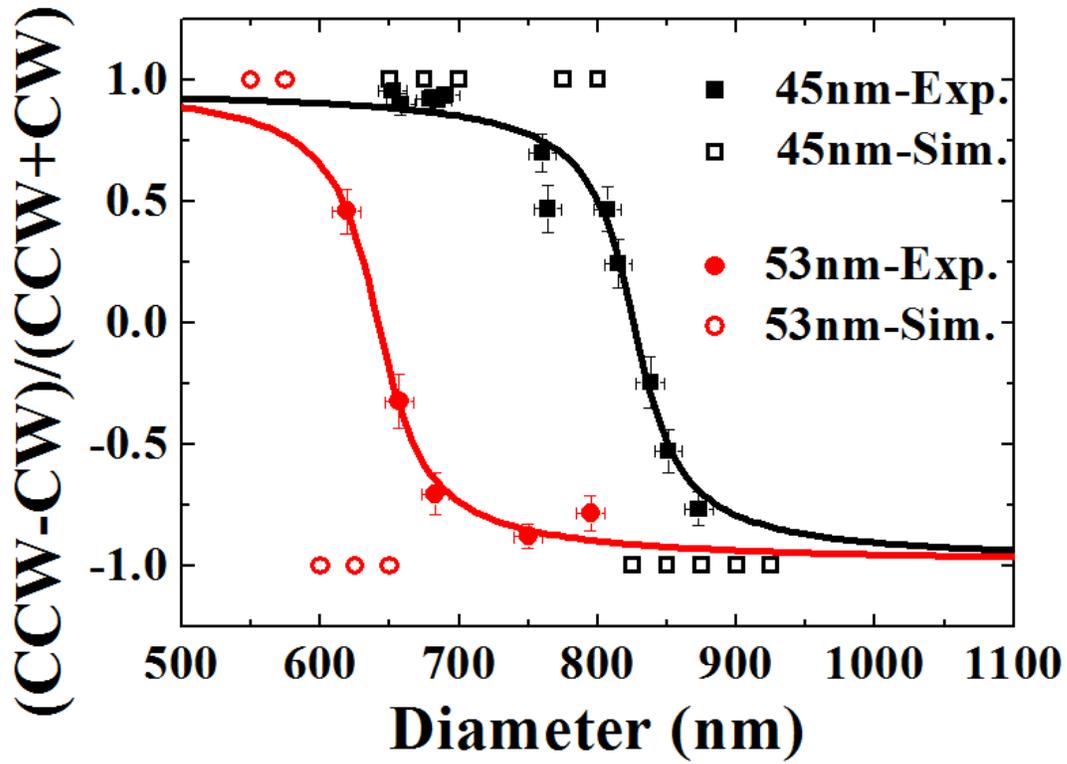

Fig. 4, Dumas et. al.